\documentclass[twocolumn,showpacs,prl,aps,amssymb]{revtex4}
\usepackage{graphicx}
\begin{document}
\title{\bf Entanglement and quantum fluctuations}

\author{ Alexander A. Klyachko and Alexander S. Shumovsky}

\address{Faculty of Science, Bilkent University, Bilkent, Ankara,
06800 Turkey}

\begin{abstract}
We discuss maximum entangled states of quantum systems in terms of
quantum fluctuations of all essential measurements responsible for
manifestation of entanglement. Namely, we consider maximum
entanglement as a property of states, for which quantum
fluctuations come to their extreme.
\end{abstract}

\maketitle
%\narrowtext

%\twocolumn

In spite of a great progress in investigation and implementation
of quantum entanglement, it still remains an enigmatic phenomenon,
which is in need of accurate definition and disclosure of
mathematical structure hidden behind it. The urgency of this is
caused by the discovery of quantum cryptography \cite{1} and
quantum teleportation \cite{2} that put entanglement at the very
heart of quantum information processing and quantum computing
(e.g., see Ref. \cite{3}).

As usually, the first thing to settle is to separate the essential
from accidental. In the case of entanglement, this touches upon
even the very definition, which still remains intuitive at great
extent. In fact, the concept of entanglement was formed under
strong influence of bipartite systems that have transparent
mathematical structure provided by the Schmidt decomposition
\cite{4}.

Probably, the definition of entanglement that the most experts are
agreeing with is as follows.

{\it "Quantum entanglement is a subtle nonlocal correlation among
the parts of a quantum system that has no classical analog. Thus,
entanglement is best characterized and quantified as a feature of
the system that cannot be created through local operations that
act on the different parts separately, or by means of classical
communication among the parts."} (See Ref. \cite{5}).

The assigning primary importance to the quantum correlations,
having no classical analog, seems to be the most essential in the
intuitive definitions of entanglement. In particular, just these
correlations are responsible for information transmission in
quantum channels \cite{6}.

At the same time, the assumption of nonlocality, which is
important for quantum communication processing, leads to a loss in
generality. In particular, this requirement is meaningless in the
case of entanglement in Bose-Einstein condensate because of the
strong overlap of wave functions of individual atoms \cite{7}.
Besides that, it leaves aside the single-particle entanglement
with respect to intrinsic degrees of freedom \cite{8,9,10}.

The entanglement is also defined in terms of violation of {\it
classical realism} described by the Bell-type inequalities
\cite{11} and Greenberger-Horne-Zeilinger (GHZ) conditions
\cite{12}. This means the violation of local classical constrains
on the correlation of measurements performed on different parts of
the system. In fact, violation of these constrains indicates the
quantum nature of states and can be observed for unentangled
states as well. An example is provided by coherent states \cite{9}
that represent an exact antithesis to entanglement.

In turn, the requirement of nonseparability of states that is
often considered as a definition, in reality is not a sufficient
condition of entanglement \cite{13}. An example of nonseparable
unentangled state is considered below.

An important property of entangled states is that one can change
the amount of entanglement by a certain operations such as the
Lorentz bust \cite{14} and SLOCC (stochastic local operations
assisted by classical communications) \cite{15,16,17}. In
particular, this means that all entangled states can be
constructed from the {\it maximum entangled} (ME) states through
the use of these operations \cite{17}. Thus, if ME states are well
defined, all other entangled states can be constructed.

The main aim of this note is to discuss a novel definition of ME
states and its consequences.

From the intuitive definitions, it follows that the maximum
entanglement should be considered as an extreme property of
quantum states maximally remote from the "classical reality"
(classical bonds on correlation of measurements). Such a
"remoteness" can be naturally specified in terms of {\it quantum
fluctuations} of observables.

As a matter of fact, the principle difference between the
classical and quantum levels of description of physical systems
consists just in the existence of quantum fluctuations
(uncertainties) in the latter case, caused by the interpretation
of  observables in terms of Hermitian operators. The {\it total
uncertainty} of all essential measurements performed over a
physical system can be used as a measure of remoteness of a state
of this system from the classical reality.

{\it We choose to interpret ME state of a given system as that,
providing the maximum remoteness from the classical reality}.

To cast this definition into a rigorous form, consider a quantum
system $S$ (not necessarily a nonlocal one) defined in the Hilbert
space $ {\mathbb{H}}(S)$. Let $\{ M_i \}$ be a set of all
essential measurements. The choice of the essential observables
depends on the physical measurements we are going to perform over
the system, or, what is the same, on the Hamiltonians, which are
accessible for the manipulation with quantum states. Let $\psi \in
{\mathbb{H}}(S)$ be a pure state. Then, the result of quantum
measurement $M_i$ is specified by the expectation value
\begin{eqnarray}
\langle M_i \rangle = \langle \psi |M_i| \psi \rangle \label{1}
\end{eqnarray}
and variance (uncertainty)
\begin{eqnarray}
V_i( \psi ) \equiv \langle (\Delta M_i)^2 \rangle = \langle
\psi|M_i^2| \psi \rangle - \langle \psi|M_i| \psi \rangle^2 ,
\label{2}
\end{eqnarray}
describing quantum fluctuations. In the case of mixed state with
the density matrix $\rho$, instead of (1) and (2) we get
\begin{eqnarray}
\langle M_i \rangle = Tr(\rho M_i), \quad V_i( \rho )=Tr[ \rho (
\Delta M_i)^2] , \label{3}
\end{eqnarray}
respectively.

We choose to specify the remoteness of quantum states from the
classical reality by the {\it total variance} that has the form
\begin{eqnarray}
{\mathbb{V}}( \psi )= \sum_i V_i( \psi ), \label{4}
\end{eqnarray}
in the case of pure states, and
\begin{eqnarray}
{\mathbb{V}}( \rho )= \sum_i V_i( \rho ), \label{5}
\end{eqnarray}
in the case of mixed states.

In the spirit of our philosophy, {\it we define ME of pure states
in ${\mathbb{H}}(S)$ by the condition}
\begin{eqnarray}
{ \mathbb{V}}( \psi_{ME} )= \max_{\psi \in { \mathbb{H}}(S)}
{\mathbb{V}}(\psi) .  \label{6}
\end{eqnarray}
In the case of mixed states, Eq. (6) is replaced by the following
\begin{eqnarray}
{ \mathbb{V}}( \rho_{ME})= \max_{\rho} { \mathbb{V}}( \rho).
\label{7}
\end{eqnarray}
This definition  means that ME states have the maximum scale of
quantum fluctuations of all essential measurements.  Eqs. (6) and
(7) represent a {\it variational principle} for maximum
entanglement similar in a sense to the maximum entropy principle
in statistical mechanics. In particular, this principle permits us
to understand how to prepare a persistent ME state, which is
necessary for teleportation and other quantum information
processes. First, we should exert influence upon the system to
achieve the state with maximum scale of quantum fluctuations.
Then, the energy of the system should be decreased up to a (local)
minimum under the condition of retention of the fluctuation scale.
The possible realization of the process in atom-photon system was
discussed in Refs. \cite{18,19}.

In a special case of interest, when the enveloping algebra of the
Lie algebra of all essential measurements ${\mathcal{L}}(M)$
contains a uniquely defined Casimir operator
\begin{eqnarray}
\hat{\mathbf{C}}= \sum_i M^2_i={ \mathbf{C}} \times \mathbb{I},
\label{8}
\end{eqnarray}
where $\mathbb{I}$ is the unit operator, the conditions (6) and
(7) can be represented in a different form. Namely, it follows
from Eqs. (2)-(5) and (8) that the maximum in (6) and (7) is
achieved if
\begin{eqnarray}
\forall i \quad \langle M_i \rangle =0. \label{9}
\end{eqnarray}
Then, the maximum total variance takes the form
\begin{eqnarray}
{ \mathbb{V}}_{max}={ \mathbf{C}}. \label{10}
\end{eqnarray}
It should be mentioned that it was observed in Ref. \cite{20} that
ME states obey the condition (9), which can be used as an
operational definition of maximum entanglement.

To illustrate the definition of ME states (4) and condition (9),
consider a system of $N$ qubits defined in the Hilbert space
\begin{eqnarray}
{ \mathbb{H}}_{2,N}=({ \mathbb{H}}_2)^N . \nonumber
\end{eqnarray}
A pure state $\psi \in { \mathbb{H}}_{2,N}$ has the form
\begin{eqnarray}
| \psi \rangle = \sum \psi_{\ell_1 \ell_2 \cdots \ell_N}
\mathbf{e}_{\ell_1} \otimes \mathbf{e}_{\ell_2} \otimes \cdots
\otimes \mathbf{e}_{\ell_N}, \label{11}
\end{eqnarray}
where $ \mathbf{e}_{\ell}=| \ell \rangle$, $\ell =0,1$, are the
base vectors in $ {\mathbb{H}}_2$. The dynamic symmetry group $G$
in the two-dimensional space ${ \mathbb{H}}_2$ is $G=SU(2)$. At
the same time, the {\it local measurements} are provided by the
Pauli operators \cite{21}
\begin{eqnarray}
\left\{ \begin{array}{ll} \sigma_1  = & |0 \rangle \langle 1+|1
\rangle \langle 0| \\ \sigma_2  = & -i(|0 \rangle \langle 1|-|1
\rangle \langle 0|) \\ \sigma_3  = & |0 \rangle \langle 0|-|1
\rangle \langle 1| \end{array} \right. \label{12}
\end{eqnarray}
that form a representation of the infinitesimal generators of the
$SL(2,{ \mathbb{C}})$ algebra ${ \mathcal{L}}^c$. The
corresponding dynamic symmetry group $G^c= \exp ({
\mathcal{L}}^c)$ is the $SL(2,{ \mathbb{C}})$ group, which is
known to be the {\it complexification} of $G=SU(2)$.

It should be stressed that the complexification of the dynamic
group plays here very important role \cite{8}. In particular,
SLOCC in an $N$-partite system is identified with the
complexification $G^c= \prod_i SL({ \mathbb{H}}_i)$ of the dynamic
symmetry group $G= \prod_i SU({ \mathbb{H}}_i)$ \cite{16,17}. The
complexification of the spin group $G=SU(2)$ also emerges as
Lorentz group, locally isomorphic to $G^c=SL(2,{ \mathbb{C}})$, in
study of relativistic transformation of entanglement \cite{14}.

The condition of maximum entanglement (9) imposes a certain
restrictions on the multidimensional matrix \cite{22} $[ \psi ]$
of coefficients in (11). Namely, the {\it parallel slices of $[
\psi ]$ should be orthogonal and have the same norm}. This result
can also be considered as the necessary and sufficient condition
of ME states \cite{8}.

Consider the three-qubit system ($N=3$). In this case, $[ \psi ]$
is a three-dimensional matrix (cube) shown in Fig. 1. The parallel
slices are represented by the faces of the cube. Consider, for
example, the parallel faces $(\psi_{000}, \psi_{010}, \psi_{100},
\psi_{110})$ and $(\psi_{001}, \psi_{011}, \psi_{101},
\psi_{111})$. The condition of orthogonality then gives the
equations
\begin{eqnarray}
Re( \psi_{000} \psi_{001}^*+ \psi_{010} \psi_{011}^*+ \psi_{110}
\psi_{111}^*+ \psi_{100} \psi_{101}^*) & =0, \nonumber \\ Im(
\psi_{000} \psi_{001}^*+ \psi_{010} \psi_{001}^*+ \psi_{110}
\psi_{111}^*+ \psi_{100} \psi_{101}^*) & =0, \nonumber
\end{eqnarray}
that coincide with the conditions $\langle \sigma_1^{(3)} \rangle
=0$ and $\langle \sigma_2^{(3)} \rangle =0$, where the superscript
indicates the measuring party. In turn, the equation
\begin{eqnarray}
|\psi_{000}|^2+|\psi_{010}|^2+|\psi_{110}|^2+|\psi_{100}|^2
\nonumber \\
=|\psi_{001}|^2+|\psi_{011}|^2+|\psi_{111}|^2+|\psi_{101}|^2,
\nonumber
\end{eqnarray}
specifying the equality of the corresponding norms, comes from the
condition $\langle \sigma_3^{(3)} \rangle =0$. All other
conditions in (9) can be considered in the same way.

It is seen that the conditions (9) in the case of $N$-qubit system
give $3 \times N$ equations. One more equation comes from the
normalization of (11). At the same time, the number of complex
elements in $[ \psi ]$ is equal to $\dim { \mathbb{H}}_{2,N}=2^N$,
which corresponds to $2 \times 2^N$ real parameters. Since
$2^{N+1} >3 \times N+1$ at $N \geq 2$, any bipartite and
multipartite system of two qubits has infinitely many ME states.
Among them, the ME states forming a basis in ${ \mathbb{H}}_{2,N}$
are important. Such a basis can be constructed in the following
way. Consider first the {\it generic} ME state of $N$ qubits
\begin{eqnarray}
\frac{1}{\sqrt{2}} (|00 \cdots 0 \rangle \pm |11 \cdots 1 \rangle
) = \frac{1}{\sqrt{2}} \sum_{ \ell =0}^1 \bigotimes_{j=1}^N | \ell
\rangle_j. \label{13}
\end{eqnarray}
It is easily seen that the two states (13) obey the condition (9).
The examples at $N=2$ and $N=3$ are provided by the Bell and GHZ
states, respectively. The basis of ME states in ${
\mathbb{H}}_{2,N}$ can be generated by the action of the {\it
local cyclic permutation operator}
\begin{eqnarray}
{ \mathcal{C}}_2= |0 \rangle \langle 1|+|1 \rangle \langle 0|
\label{14}
\end{eqnarray}
on the generic states (13) $(2^{N-1}-1)$ times (also see Ref.
\cite{2}). Here the operator (14) formally coincides with
$\sigma_1$ in (12). In the case of qudits ($d$ degrees of freedom
per party), this operator takes the form
\begin{eqnarray}
{ \mathcal{C}}_d=|0 \rangle \langle 1|+ \cdots +|d-1 \rangle
\langle 0| \nonumber
\end{eqnarray}
so that ${ \mathcal{C}}^d={ \mathbb{I}}$.

Consider now a few important corollaries of the above definition
of ME states. \\ $\bullet$ {\it Corollary 1}. The classification
of quantum states according to the scale of quantum fluctuations
is known since the discussion of {\it coherent states} (e.g., see
\cite{23,24}). The coherent states have the minimum scale of
quantum fluctuations and therefore they are considered as almost
classical states. Thus, by definition, ME states represent an
exact antithesis to coherent states. \\ $\bullet$ {\it Corollary
2}. The definition of ME states (6) or (7) is independent on
whether the system is nonlocal or not. For example, it can be used
to specify ME states of a particle with respect to internal
degrees of freedom. An example is provided by $\pi$-mesons that
are built from the up $u$ and down $d$ quarks as follows \cite{25}
\begin{eqnarray}
\pi^+ \rightarrow u\bar{d}, \quad \pi^- \rightarrow \bar{u}d,
\quad \pi_0 \rightarrow \frac{u\bar{u}-d\bar{d}}{\sqrt{2}} .
\nonumber
\end{eqnarray}
The first two particles $\pi^{\pm}$ correspond to the coherent
states of internal (quark) degrees of freedom, while $\pi^0$ is
specified by ME state with respect to quarks. Since these two
states have quite different levels of quantum fluctuations,
$\pi^0$ meson should be less stable than $\pi^{\pm}$. In fact, the
ratio of lifetimes is $\tau_0/ \tau_{\pm} \sim 3 \times 10^{-9}$.
The single-particle entangled states are possible if the number of
internal degrees of freedom exceeds two (single qutrit etc.).
\\ $\bullet$ {\it Corollary 3}. The definition of ME states (6)
can also be applied to the photon field, when the dynamic symmetry
is specified by the Weyl-Heisenberg group. Choosing the
measurements as the field quadratures
\begin{eqnarray}
q= \frac{1}{2} (a+a^+), \quad p=\frac{-i}{2} (a-a^+), \quad
[a,a^+]=1 \nonumber
\end{eqnarray}
and employing condition (9), it is easily seen that the Fock
number state $|n \rangle$ and squeezed vacuum state give $\langle
q \rangle = \langle p \rangle =0$ and provide the total variances
\begin{eqnarray}
{\mathbb{V}}_{Fock}=(2n+1)/2, \quad { \mathbb{V}}_{squeez}=(2
\cosh r -1)/2, \nonumber
\end{eqnarray}
respectively. Here $r$ is the squeezing parameter. A certain
difference between the variances is caused by the fact that the
Weyl-Heisenberg algebra has no uniquely defined Casimir operator.
The number and squeezed vacuum state represents a kind of {\it
parametric} ME states because the scale of quantum fluctuations is
specified by the parameters $n$ and $r$, respectively. The real ME
states are achieved in the limit $n,r \rightarrow \infty$.

Of course, some other entangled states of light, specified by the
measurements belonging to the $SU(N)$ subalgebras in the
Weyl-Heisenberg algebra and corresponding to the polarization and
angular moment of photons, can also be considered in terms of
definition (6) and (7).
\\ $\bullet$ {\it Corollary 4}. The total variance (4) cannot be used as a measure of
entanglement. The unentangled states may manifest even more
quantum fluctuations than entangled states but not ME states. For
example, the three-qubit $W$ state \cite{16} has ${
\mathbb{V}}_W=8+2/3$. The maximum entangled GHZ state has ${
\mathbb{V}}_{GHZ}=9$, so that the level of quantum fluctuations in
$W$ state is quite close to maximum. At the same time, $W$ state
is not entangled because the 3-tangle \cite{26}, which is similar
to Cayley's hyperdeterminant \cite{27} and is an entangled
monotone in the case of three qubits, is equal to zero. Thus,
nonseparable $W$ state is not entangled. In turn, the state
\begin{eqnarray}
|\psi_x \rangle = x(|000 \rangle +|111 \rangle )+y(|001 \rangle
+|110 \rangle ), \label{15}
\end{eqnarray}
where $x^2+y^2=1/2$, is entangled but not maximum entangled at $x
\neq 1/2, 1/ \sqrt{2}$. In the intervals $0.122<x<0.5$ and
$0.5<x<0.696$, we get
${ \mathbb{V}}_x<{ \mathbb{V}}_W$. \\
$\bullet$ {\it Corollary 5}. The definition (6) and possibility to
construct the entangled states from ME states by means of SLOCC
put the notion of entanglement within the framework of geometric
invariant theory \cite{8} (concerning geometric invariant theory,
see Ref. \cite{28}). This permits us to introduce a new {\it
universal measure} of entanglement which is the length of minimal
vector in complex orbit \cite{8}
\begin{eqnarray}
\mu(\psi)= \min_{g \in G^c} |g \psi |^2. \label{16}
\end{eqnarray}
Here $\psi \in { \mathbb{H}}(S)$ is a state of quantum system $S$
and $G^c$ is the complexified dynamic symmetry group in ${
\mathbb{H}}(S)$. This measure represents an entangled monotone,
which is equal to zero for unentangled states and achieves maximum
for ME states. The measure (16) coincides with the concurrence
(determinant of $[ \psi ]$ in (11) at $N=2$) in the case of two
qubits and with the square root of 3-tangle for three qubits. It
can also be calculated in the case of four qubits (all geometric
invariants in four-qubit system were calculated in Ref.
\cite{29}). The description of entanglement and its proper measure
(not necessarily of maximum entanglement) within the framework of
geometric invariant theory deserves a special, more detailed
discussion.

Summarizing, we should stress that the definition of ME states in
terms of extreme quantum fluctuations agrees with that based on
the consideration of quantum correlations, whose maximum in many
cases can be expressed in terms of condition (9) as well. At the
same time, definition in terms of the variational principle (6) or
(7) is more general and has an evident heuristic advantage. It
lays bare the physical essence of the phenomenon as the
manifestation of quantum fluctuations at their extreme and reveals
the mathematical structure hidden behind the entanglement.

The most of examples in this note were referred to pure states.
They can be easily generalized on the case of mixed states because
the latter can be treated as a pure state of a certain doublet,
consisting of the system $S$ and its "mirror image" \cite{30}.

\begin{figure}
\includegraphics[scale=0.5]{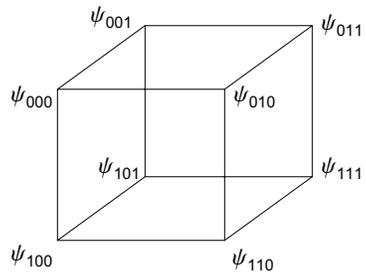}
\caption{Structure of the three-dimensional matrix $[ \psi ]$ of
three qubits. Vertices of the cube are associated with the
coefficients $\psi_{ijk}$ in Eq. (11) at $N=3$.}
\end{figure}

\end{document}